# Gas-liquid Nucleation at Large Metastability

Mantu Santra, Rakesh S. Singh and Biman Bagchi[*]

Solid State and Structural Chemistry Unit,

Indian Institute of Science, Bangalore 560012.

[*]Email: bbagchi@sscu.iisc.ernet.in

## *Abstract*

Nucleation at *large metastability* is still largely an unsolved problem, although is a problem of tremendous current interest, with wide practical value. It is well-accepted that the classical nucleation theory (CNT) fails to provide a qualitative picture and gives incorrect quantitative values for such quantities as activation free energy barrier and supersaturation dependence of nucleation rate, especially at large metastability. In this article, we present a powerful alternative formalism to treat nucleation at large supersaturation. This formalism goes over to the classical picture at small supersaturation where CNT is expected to be valid. The new theory is based on an extended set of order parameters in terms of *k-th* largest liquid-like clusters where $k=1$ is the largest cluster in the system, $k=2$ is the second largest cluster and so on. We derive an analytic expression for the free energy of formation of the *k-th* largest cluster which shows that at large metastability the barrier of growth for the few largest liquid-like clusters disappear, the nucleation becomes collective and the approach to the critical size occurs by barrierless diffusion in the cluster size space. The expression for the rate of barrier crossing predicts a weaker supersaturation dependence than that of CNT at large metastability. Such a cross-over behavior has indeed been observed in recent experiments but eluded an explanation till now. In order to understand the large numerical difference between simulation predictions and experimental results, we carried out a study of the dependence on the range of intermolecular interaction of both the surface tension of an equilibrium planar gas-liquid interface and the free energy barrier of nucleation. Both are found to depend significantly on the range of interaction for a Lennard-Jones potential, both in two and three dimensions.

## I. Introduction

First order phase transitions usually occur via the formation of a droplet of the stable phase within the metastable bulk phase through an activated process called nucleation **[1-9]**. The classical nucleation theory (CNT) (of Becker-Döring-Zeldovich) provides a simple yet elegant description of homogeneous nucleation in terms of this free energy barrier, with size of the cluster as the sole order parameter describing nucleation. CNT assumes that a spherical droplet of the new stable phase grows in a sea of parent metastable bulk phase by addition of single molecules and this droplet has to grow beyond a certain ''critical size'' ($R^*$) to compensate for the energy required to form the surface between the two phases. The free energy of formation of a droplet of radius $R$ is

$$\Delta G(R) = -\frac{4\pi}{3} R^3 |\Delta G_V| + 4\pi R^2 \gamma, \quad (1)$$

where $|\Delta G_V|$ is the free energy difference per unit volume between the daughter and the parent phases and $\gamma$ is the surface tension of the interface between them. The above relation gives the following expressions for the size ($R^*$) of the critical nucleus and the free energy barrier ($\Delta G(R^*)$)

$$R^* = \frac{2\gamma}{\Delta G_V}$$
$$\Delta G(R^*) = \frac{16\pi \gamma^3}{3(\Delta G_V)^2} \quad . \quad (2)$$

Note that the free energy barrier depends more strongly on the surface tension than on free energy difference. From the above we get the expected result that both the critical cluster size ($R^*$) and the free energy barrier ($\Delta G(R^*)$) decrease with increase in supersaturation or supercooling and the rate of nucleation increases. While several aspects of this classical nucleation theory have recently been analyzed critically **[10-16]** and different lacunae have been removed, yet the important problem of the mechanism of nucleation at large supercooling, especially that near the gas-liquid spinodal, remains unresolved and somewhat controversial. The well-known droplet model **[17]** suggests a *kinetic origin* of the spinodal. Mean field theoretic investigations of the spinodal nucleation close to the critical point predicts that the critical size should diverge with increasing quench depth and the process can have 'universal characteristics' **[14, 18]**. For the liquid to vapour transition, the critical vapour nucleus has been found to be a large weblike nonspherical cavity for a wide range of superheating suggesting a strong deviation from the CNT **[19]**.

*Nucleation at large metastability is an important issue in many branches of physics and chemistry, starting from astrophysics to materials chemistry.* The issues regarding surface tension and nucleation are now being intensely debated in the area of glass transition at low temperature liquids **[20-22]** and also in protein folding **[23]**. In the case of glass transition, an unusual $1/\sqrt{R}$ dependence of surface tension of the size $R$ of the nucleus has been used to explain, rather satisfactorily, many aspects of glass transition behaviour. Theoretically, the justification of $1/\sqrt{R}$ dependence comes from random Ising model **[24]**. In the case of gas-liquid nucleation or liquid-crystal nucleation or in synthesis of solids from metastable sols, we do not have quantitative understanding of the dependence of surface tension on the metastability or the size or the shape of the cluster and such dependencies can be non-trivial.

There are several issues regarding nucleation at large metastability. First of course is the issue of the quantitative accuracy. It is now well-established that the use of CNT with free energy from thermodynamic data and surface tension obtained from the equilibrium gas-liquid coexistence with a planar interface do not provide quantitative agreement. The second issue concerns the validity of the free energy decomposition embodied in **Eq. (1)**. While such decomposition can be valid at low supersaturation when the size of the critical cluster is large, it becomes questionable when the embryo size becomes small at large supersaturation. Density functional theoretical approach would suggest the calculation of the free energy barrier of a critical cluster directly from the density profile **[25-27]**. This approach allows the embryo to have characteristics different from the final product state. However, DFT has been applied till now only to spherical nucleus. In addition, it is an equilibrium theory. A dynamical density functional theory along the approach initiated by Oxtoby and co-workers **[28]** is yet to be developed.

Bustillos *et al*. **[29]** have constructed the free energy surface for the three-dimensional Ising model as a function of the magnetization and the external field, and found that for states deep in the two-phase region (where the system was expected to be mechanically unstable) the nucleation barrier does become small *but does not disappear,* suggesting that *'there is no spinodal'*! Very recently Parrinello and co-workers have studied the freezing of a Lennard-Jones (LJ) fluid as a function of the degree of supercooling to find that the nucleation acquires a spinodal character for deeper quenches and crystallization proceeds in a more continuous and collective fashion and becomes spatially more diffuse as the spinodal is approached **[30]**. That is, several large clusters grow at large metastability and the ultimate fate of these clusters is stochastic. This prediction has also been confirmed in gas-liquid nucleation **[31-33]**. At large metastability many clusters grow simultaneously and the CNT description in terms of the growth of a single cluster does not seem to hold at all.

Therefore, one needs to develop a theoretical formalism that can describe the simultaneous growth of several clusters. We then need to introduce a set of order parameters that allows us to order the clusters according to their sizes. We have developed such formalism **[32]**. We calculate the free energy of formation of the *k-th* largest cluster of size *n* under given thermodynamic conditions. In this notation, *k*=1 stands for the largest cluster in the system, *k*=2 is the second largest cluster, *k*=3 the third largest cluster and so on. We find a relation between free energy of formation of *k-th* largest cluster of size *n*, $F_k^L(n)$ and the CNT free energy, *F(n)*.

The organization of the rest of the paper is as follows. In the next section, we first construct a two dimensional free energy surface as an extension of the CNT free energy surface and demonstrate that such an extended description can provide useful insight into nucleation mechanism at large metastability. In section III we introduce the larger set of order parameters and evaluate the free energy of formation of *k-th* largest cluster. In section IV we study the crossover mechanism that emerges at large metastability. This section also includes the mean first passage time calculation of the rate. In section V we discuss the gas-liquid nucleation in two dimensional Lennard-Jones systems. In section VI we briefly discuss the effects of interaction range on nucleation. Section VII concludes with a summary.

## II. Two Dimensional Free Energy Surface: Elucidation of Nucleation near Spinodal

In this section we discuss spinodal nucleation by first removing the restriction of the existence and growth of a single nucleating cluster and secondly, by calculating free energy as a function of two relevant order parameters for Lennard-Jones fluid. The two reaction coordinates are the total number of particles present in the system (grand canonical (μVT) ensemble) and the 'liquidness' (analogous to the magnetization in the Ising model) of the system. The latter is given by the total number of *liquid-like particles* ($N_{Liq}$) identified by its local density. Following the definition by Stillinger [34] and the successful modification by ten Wolde and Frenkel [13] a particle is considered to be liquid-like if it has more than four nearest neighbours within a cut-off distance ($R_c=1.5\sigma$). Liquid-like particles that are connected by neighbourhood (within the cut-off distance) form *liquid-like clusters*. **Fig. 1(a)** and **1(b)** show the calculated free energy surface of formation of liquid-like clusters in a system of Lennard-Jones spheres at two different supersaturations S (given by $P/P_0$, where P is the pressure of the system and $P_0$ is the pressure at coexistence at the same temperature) equal to 1.8 and 2.4. With the interaction range truncated at 2.5 σ (where σ is the usual Lennard-Jones diameter), the spinodal point is estimated to be between 2.5-2.6. It is to be noted that this estimate is in close agreement with those of Moody *et al.* [35] (estimate it to be 2.7 from surface tension calculations). At intermediate supersaturation (S~1.8), still away from the spinodal point, both the activation barrier and the number of liquid-like particles at the barrier are large (**Fig. 1a**). The latter is about 50 (we discuss later that the critical nucleus alone contains nearly all the liquid-like particles) and the activation energy is about 9.5 $k_BT$. On the other hand, near the spinodal (S~2.4), the free energy surface near the saddle is *very flat* (**Fig. 1(b)**). Here the number of liquid-like particles at the saddle is just about 35 and the free energy barrier from the minimum is less than 4 $k_BT$. Importantly, these liquid-like particles are dispersed among several intermediate sized clusters, as discussed below in more detail.

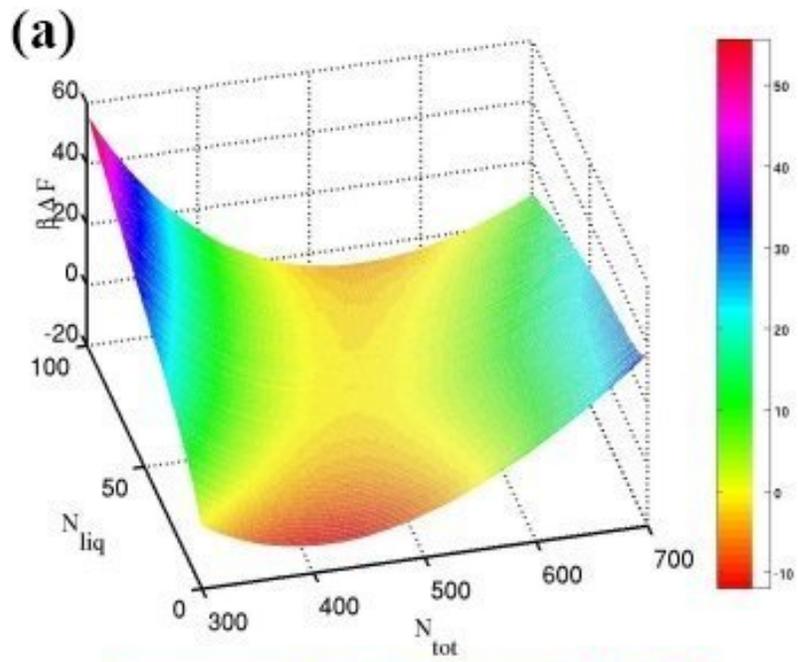
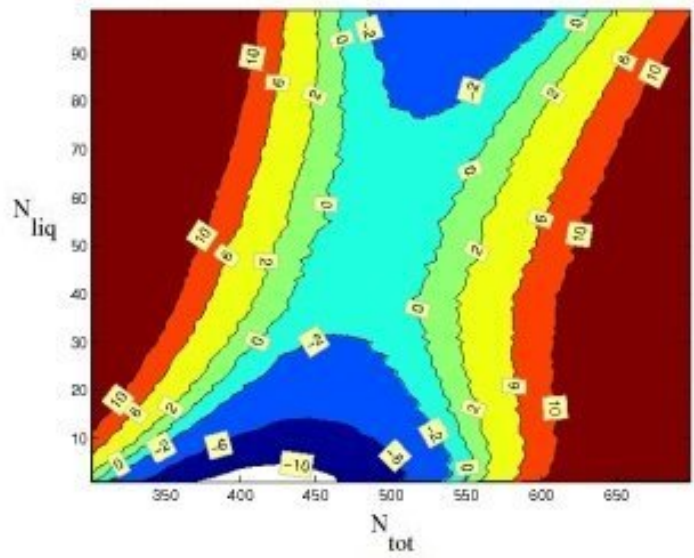

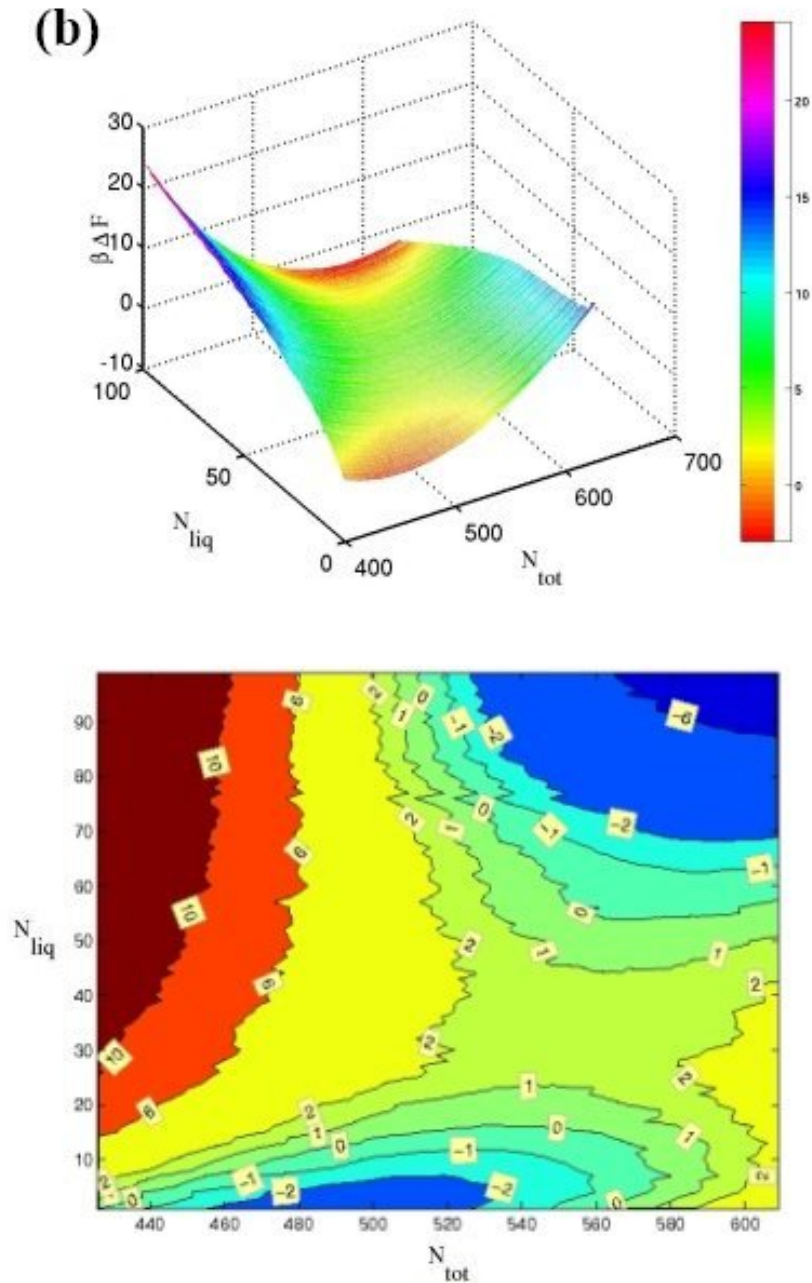

**Figure – 1.** The computed three dimensional free energy surfaces and contour plots computed in grand canonical (μVT) ensemble at reduced temperature $T^*=0.741$ and volume $V = (25\ \sigma)^3$. Activity is defined as $\xi=\exp(\mu/k_BT)/\lambda^3$, where $\lambda$ is the de Broglie wave length and $\mu$ is the chemical potential. (a) 3-dimensional free energy surface and (bottom) 2-dimensional contour diagram for $\xi=0.018$, which approximately corresponds to supersaturation S~1.8 in the NPT simulation. This supersaturation is large but still may be described semi-quantitatively by the CNT theory. (b) At a larger supersaturation S~2.4 (close to the spinodal) and $\xi=0.020$. These figures have been computed

**with a cut-off and shift of the range of the interaction potential at $r_{cut}$ = 2.5 σ. When the range of the interaction potential is increased (as discussed later), the largest metastability that can be studied by simulation increases significantly but the picture at the largest metastability remains essentially the same [31].**

In the cluster size distribution of liquid-like particles, computed at two different regions of the free energy surface, one near the saddle region and one in the metastable minimum, it is observed that the system at the saddle has a relative abundance of intermediate size clusters as compared to the metastable gas phase. This difference becomes particularly important as spinodal is approached. In contrast, at low supersaturation, the critical cluster contains all of the liquid-like particles and the homogeneous nucleation assumption appears to be valid.

Unlike chemical reactions, in nucleation, the activation of the largest cluster controls the phase transition of the entire system. Once the largest cluster crosses the critical size it grows rapidly. By that time any other cluster reaches the critical size it engulfs the entire system leading to the phase transition. So, it is more appropriate to consider the largest cluster as the reaction coordinate (or order parameter). **Fig. 2** shows the computed free energy surfaces of largest liquid like cluster for a wide range of supersaturations across the spinodal. At relatively small supersaturation (S=1.53) the classical picture prevails unambiguously. But as the supersaturation is increased, both the size of the critical cluster and the barrier height become progressively smaller and *a minimum appears at sub-critical cluster size* ($N_{largest}$~5)! At S=2.4 this minimum is very pronounced and essentially responsible for the existence of the nucleation barrier. As the supersaturation is increased towards the spinodal, *the free energy barrier becomes lower* and the minimum at intermediate cluster size becomes deeper and shifts to larger size. Finally it reaches an inflection point at the spinodal beyond which the expected barrier-less continuous growth takes over. The appearance of the free energy minimum at small liquid-like cluster size indicates the relative abundance and importance of such clusters in the system near the spinodal. Thus, as the degree of supercooling or super-saturation is increased, the growth of the stable phase is not through a single critical nucleus any more and becomes spatially diffuse. The system attains a spinodal like character and responds more collectively and *the importance of a critical nucleus is diminished*. This view, as discussed later, is also justified through non-equilibrium quench simulations.

The above picture remains essentially same in Ising models also. This confirms that the above observations are rather general. But we must note that the nucleation mechanism can notably differ in these two models since in Ising model the growth proceeds by single spin flips and the clusters can not physically move as compared to the liquid-like clusters.

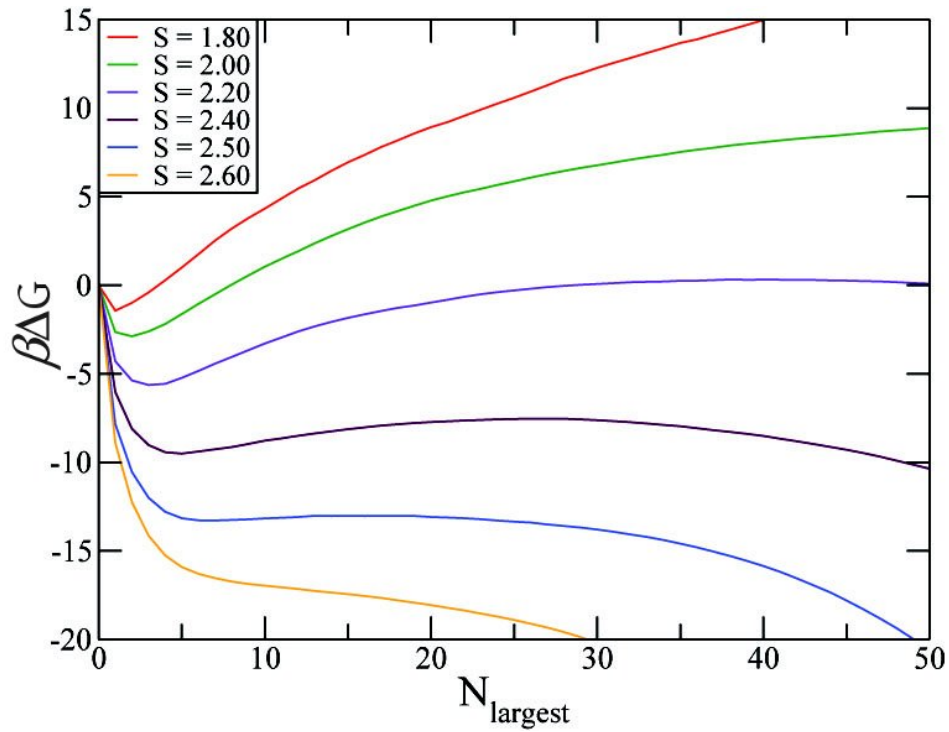

**Figure – 2. Free energy versus size of the largest liquid-like cluster in the NPT ensemble ($T^*=0.741$) for different supersaturations (S) across the spinodal in LJ system [31].**

The abundance of the small sized liquid-like clusters suggests the possibility of their coalescence as a mechanism of the formation of the liquid phase. Equilibrium molecular dynamics simulations that consisted of a temperature quench at a constant volume can reveal this mechanism. **Fig. 3** shows the difference between the growth of the largest cluster and the total number of liquid-like particles. There are several remarkable features in these growth trajectories near the spinodal.

(i) *The separation between the largest cluster size and total number of liquid-like particles grows enormously as the supersaturation is increased*. That signifies formation of liquid-like particles throughout the system apart from the largest cluster.

(ii) Second, at S~2.4, the trajectory shows severe fluctuations due to the coalescence and break-ups involving the largest cluster. For even higher supersaturations these jumps are more pronounced and the largest cluster can break-up even beyond the critical size. For vivid picture we include relevant snapshots of the system at different *S* in **Fig. 4**. This essentially supports the view of the emergence of a spatially diffuse collective mechanism for growth of the stable phase at very high supersaturation.

(iii) Coalescence as a vehicle of cluster growth is not significant for cluster sizes below the critical size. Rather, this mechanism becomes increasingly operative once the clusters cross the critical size. This is also clear from **Fig. 3**.

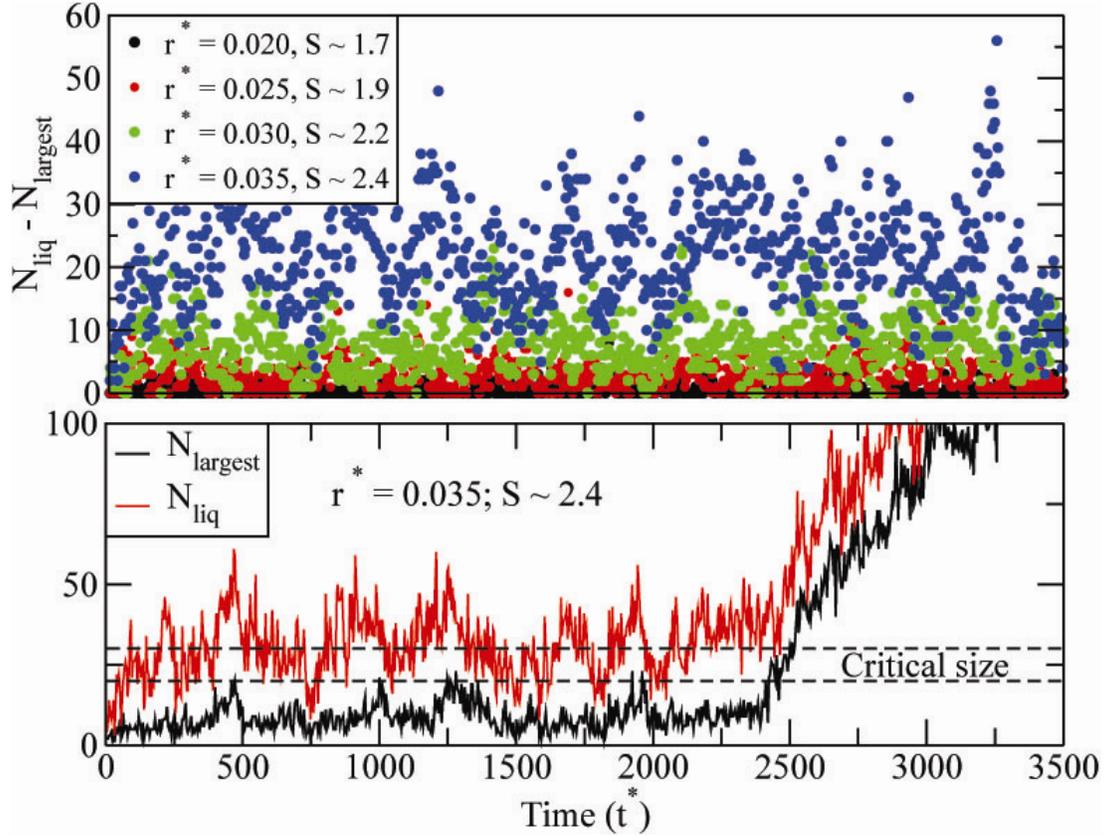

**Figure – 3. (top)** Evolution of the difference between the total number of liquid-like particles ($N_{liq}$) and the size of the largest liquid-like cluster ($N_{largest}$) subsequent to a temperature quench (from $T^*=1.0$ to $T^*=0.741$). Note that at very high supersaturation the separation between $N_{liq}$ and $N_{largest}$ grows markedly. **(bottom)** Time evolution of both $N_{liq}$ and $N_{largest}$ is shown for S~2.4. It shows huge fluctuations in both the quantities indicating coalescence of the largest cluster with smaller clusters and break-ups. Note that the largest cluster can break-up even beyond the critical size [31].

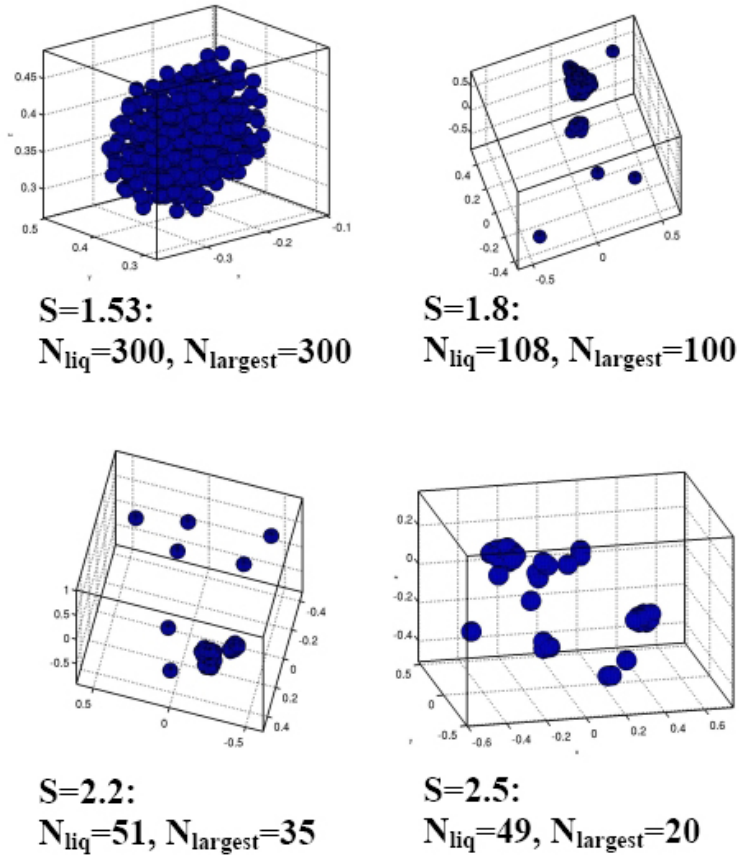

**Figure – 4. Snapshots of the system at four different supersaturations (S): They show all the liquid-like particles of the system while the largest cluster is the critical cluster. For higher supersaturations, multiple large clusters are forming around the critical cluster and growth of the liquid phase becomes spread over the whole system rather than the single 'critical cluster' [31].**

(iv) Previous analyses of the clusters near the critical point of Ising model suggest that the clusters become ramified near the spinodal **[35]**. In contrast, we find the clusters to be relatively compact for LJ fluid even at S=2.4. Theoretical studies on spinodal nucleation near the gas-liquid critical point predicted the existence of a diverging length scale. The present study at lower temperatures finds no evidence of any such diverging length. However, we do find the emergence of a large number of liquid-like, sub-critical clusters indicating flatness of the free energy surface which is also a hallmark of all critical phenomena.

(v) It is the appearance of intermediate size liquid-like particles which may render spinodal decomposition a critical phenomenon like character, even at large supercooling. That is, the growth spreads all over the system and becomes spontaneous. As already mentioned, the growth also becomes barrierless.

(vi) In accord with the studies of Parrinello and co-workers **[30]** we reiterate that while CNT remains valid for low to moderate supersaturation, it breaks down at large supersaturation or supercooling. The stable phase starts forming in a spatially diffuse, collective, and more continuous fashion. The role of surface tension (if any) has remained unclear.

(vii) As discussed later, kinetics of phase transition depends on the range of interaction potential. The convergence is found to be rather slow, and one needs to include interaction potential up to a range of $\sim 7\sigma$ to obtain the converge result. As the range is increased from the typical value of $2.5\ \sigma$ employed in many simulations, the value of the surface tension increases rather sharply, and the liquid state also becomes more stable. That is, both $\gamma$ and $\Delta G_V$ increases as the range of the interaction potential increases. However, the basic conclusions, like the emergence of the collective behaviour and the barrierless growth of largest clusters all remain unchanged.

## III. Free Energy of $k^{th}$ Largest Clusters

In the above section we have discussed that significant new insight can be gained by studying the growth and nucleation not just as a function of one order parameter (the size of the cluster) as in CNT, but as a function of several relevant order parameters. Such a need clearly arises because as we approach the spinodal limit, the growth becomes more collective and diffuse, and one may fail to obtain the correct picture by looking the growth of one cluster only. We have therefore looked at the nucleation scenario as a function of many cluster sizes, by evaluating free energy and by following such growth in simulations.

Recently, using order statistics **[36]** an expression for the free energy of the *k-th* largest cluster of size *n* has been obtained and is given by **[32]**,

$$\beta F_k^L(n) = \beta F(n) - \ln\left\{\frac{M!}{(k-1)!(M-k)!}\right\} - (M-k)\ln\{G(n)\} \\ -(k-1)\ln\{1-G(n)\} + \ln(M_n)$$ 
(3)

where $\beta F(n) = -\ln P(n)$ is the free energy of a cluster of size *n* and *P(n)* is the probability density function of *n*. $M_n$ is the average number of clusters of size *n*. *M* is the average number of total clusters present in the system and is the summation of $M_n$ over *n*. *M* includes both the liquid-like clusters and the single gas particles. *G(n)* is the cumulative probability distribution function of *P(n)*. The only approximation made in deriving **Eq. (3)** is that the clusters are non-interacting.

By putting *k* = 1 in **Eq. (3)** we obtain the free energy of largest cluster as,

$$\beta F_1^L(n) = \beta F(n) - \ln(M) + \ln(M_n) - (M-1)\ln\{G(n)\}$$ 
(4)

In simulations, it has been shown that the free energy barrier of the largest cluster, $F_1^L(n)$, is volume dependent and given approximately by $\beta \Delta F_1^L = \beta \Delta F - \ln V$ [**37, 38**].

The occurrence of the ln$V$ term can be easily rationalized in the following way. Since $M$ is the total number of clusters in the system, and since the density of the gas is still quite small, one can use Mayer's theory of condensation to obtain the relationship between $M$ and the total volume $V$ as $PV = Mk_BT$. Non-trivial part of volume dependency enters through the last term of **Eq. (4)**. $\beta F_1^L(n)$ is limited by the inclusion of integral probability of all the small clusters present in the system. Therefore, as the system size is increased the probability of a particular size of the largest cluster will involve more number of clusters' probability integral *i.e.* the last term in **Eq. (4)** will increase. As a result the pre-critical minimum in the $\beta F_1^L(n)$ will move towards larger value as well as the barrier $\beta \Delta F_1^L$ will decrease.

**Fig. 5(a)** depicts the free energy surfaces, $F_k^L(n)$ obtained from **Eq. (3)**, using the simulation data of the free energy $F(n)$. The figure shows that there is an interesting progression of $F_k^L(n)$ to $F(n)$, as $k$ is increased from $k = 1$ to larger value of $k$. Nearly identical $F_k^L(n)$ is obtained from simulations, as shown in **Fig. 5(b)**.

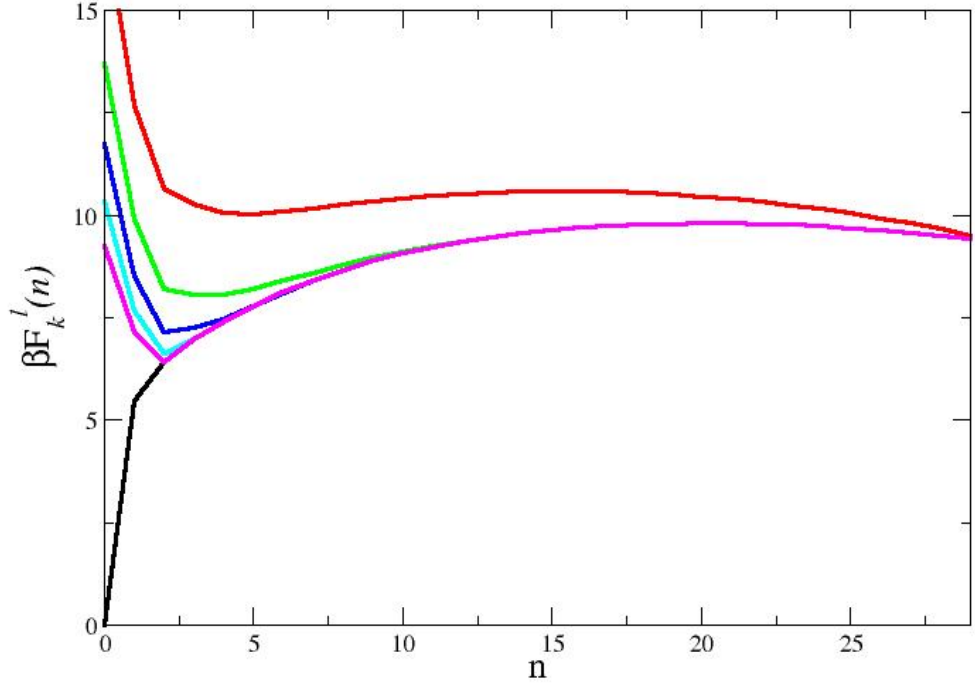

**Figure - 5(a).** Free energy of the first five largest clusters calculated from eqn. (3) at $\rho^*$=0.04158 and $T^*$=0.741 for $N$=1000. Red line is the free energy of the largest cluster ($k$ = 1), green is for $2^{nd}$ largest ($k$ = 2), blue is for $3^{rd}$ largest ($k$ = 3), cyan is for $4^{th}$ largest ($k$ = 4), magenta is for $5^{th}$ largest ($k$ = 5) cluster. The black line shows the free energy of a cluster obtained from simulation [32].

The minima at pre-critical sizes in **Fig. 5(a)** and **Fig. 5(b)** indicate that during equilibration of the system many small clusters are formed spontaneously having size much less than critical cluster size. It also suggests that the nucleation is a two step process: initial spontaneous growth of few clusters up to certain pre-critical sizes followed by an activated barrier crossing process **[30]**. The progression of $F_k^L(n)$ to $F(n)$ clearly shows that the clusters with larger $k$ value face larger barrier.

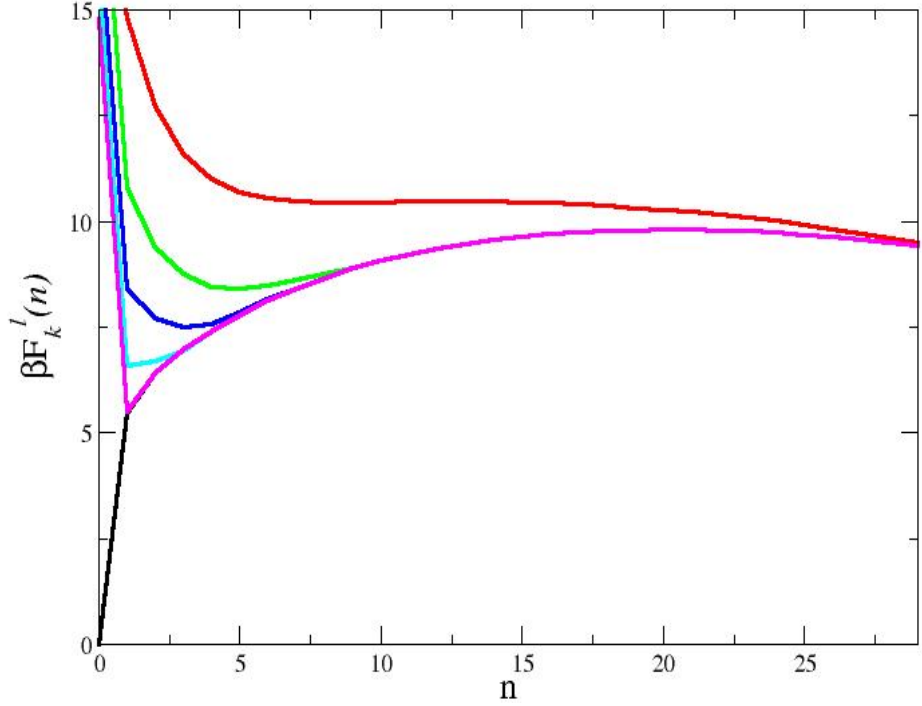

**Figure – 5(b). Free energy of the first five largest clusters, $F_k^l(n)$, with $k$=1 to 5, computed in simulation at $\rho^*$=0.04158, $T^*$ =0.741 and $N$=1000. Red line is the free energy of the largest cluster ($k$ = 1), green is for $2^{nd}$ largest ($k$ = 2), blue is for $3^{rd}$ largest ($k$ = 3), cyan is for $4^{th}$ largest ($k$ = 4), magenta is for $5^{th}$ largest ($k$ = 5) cluster. The black line shows the free energy of a cluster obtained from simulation [32].**

## IV. Crossover Behavior and Mean First Passage Time (MFPT)

As long as the largest cluster has barrier height significantly larger than $k_B T$, it is only the largest cluster that alone crosses the barrier, and the others follow at a much greater interval. See **Fig. 6(a)** for illustration of such an event. By the time the second cluster could cross the barrier, the largest cluster has already engulfed the entire system. Thus, at low supersaturation, the basic hypothesis of the classical nucleation theory remains valid. However, the situation undergoes a complete change at high supersaturation. At large $S$, many clusters now cross the nucleation barrier as the free energy barrier of these large clusters disappears. At this point $2^{nd}$, $3^{rd}$,… largest clusters all face barrier height of the order of $k_BT$. So, at this point, the time scale of crossing the barrier for many *k-th* largest clusters become comparable to each other and they start growing competitively. There is no clear separation of time scale between nucleation and growth. **Fig. 6(b)** shows such a barrierless growth.

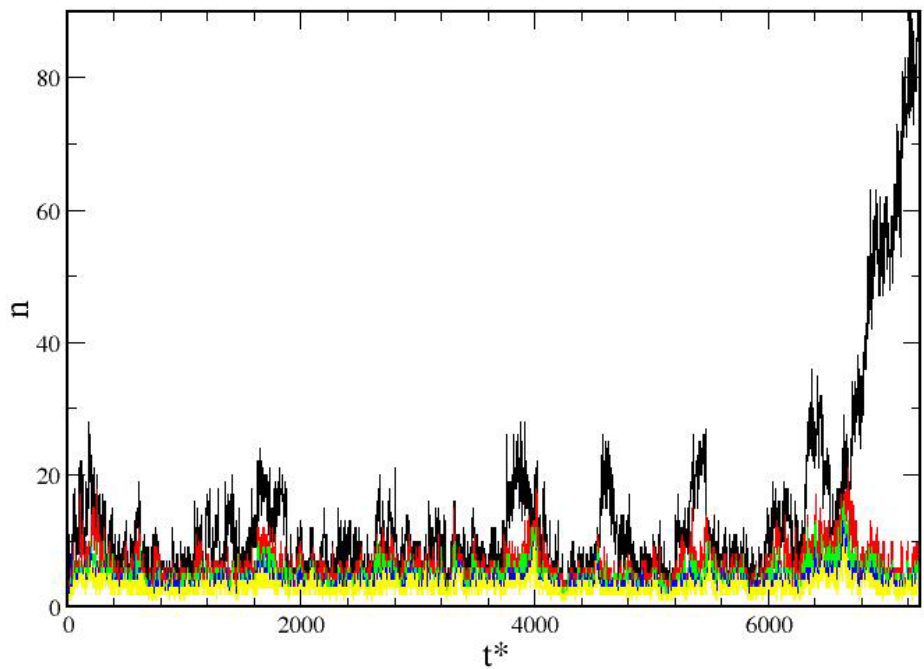

**Figure – 6(a). Growth of the $1^{st}$ five largest clusters at $\rho^*$=0.0334083 and $T^*$=0.741 for $N$=4000. Black line is for the largest cluster ($k = 1$), red is for $2^{nd}$ largest ($k = 2$), green is for $3^{rd}$ largest ($k = 3$), blue is for $4^{th}$ largest ($k = 4$), yellow is for $5^{th}$ largest ($k = 5$) cluster. The figure shows the barrier crossing of only one cluster (largest cluster) through activated process [32].**

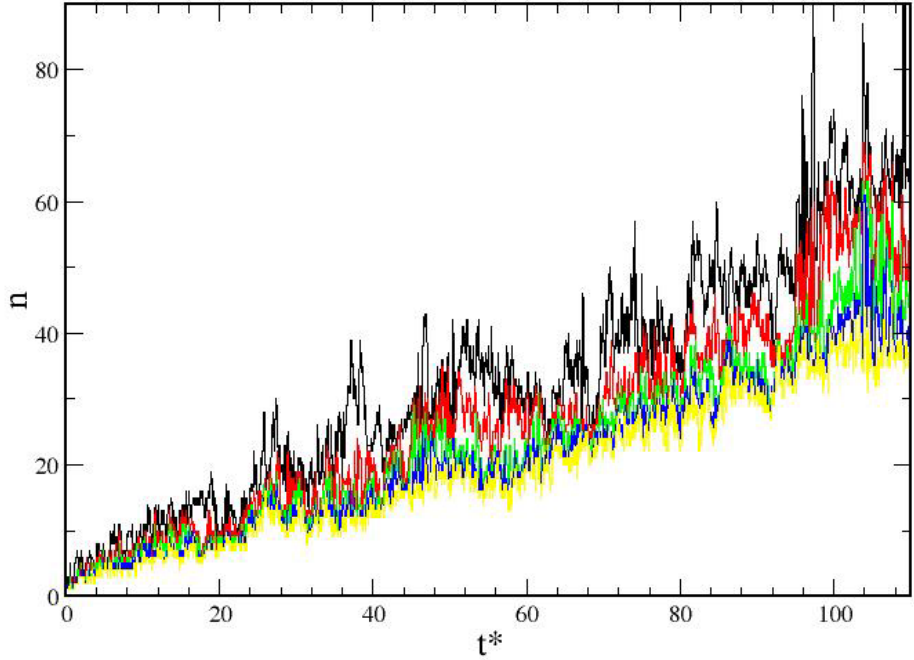

**Figure – 6(b). Growth of the $1^{st}$ five largest clusters at $\rho^*=0.05545$, $T^*=0.741$ for $N=4000$. Black line is for the largest cluster ($k = 1$), red is for $2^{nd}$ largest ($k = 2$), green is for $3^{rd}$ largest ($k = 3$), blue is for $4^{th}$ largest ($k = 4$), yellow is for $5^{th}$ largest ($k = 5$) cluster. The figure shows multiple cluster growth through barrierless diffusion [32].**

In experiments **[39-43],** the rate of nucleation, $J$, is measured by counting the number of clusters, which have crossed the barrier, per unit time in unit volume of the system,

$$J = \frac{N_d}{\tau_{exp} V}. \qquad (5)$$

Here $\tau_{exp}$ is the experimental observation time, $N_d$ is the number of clusters which have crossed the barrier within the time $\tau_{exp}$, and $V$ is the volume of the system. In many computer simulation studies a special case of **Eq. (5)** is considered where $N_d = 1$, that is, only one cluster grows (or assumed to grow). At large supersaturation $N_d$ becomes much larger than unity and above approximation ($N_d = 1$) breaks down.

Additionally, *all the theoretical formulations of nucleation rate invoke a steady state assumption. In experiments, on the contrary, one measures only the initial rate*! So, it is important to analyze the difference between the steady state and the initial rate. This initial rate can be obtained from the mean first passage time (MFPT) of a liquid-like cluster. Therefore, the free energy barrier and the corresponding barrier crossing time of the largest cluster can give a useful insight on this unresolved large discrepancy between theory and experiment. We derive an expression of the MFPT a bit later.

The change of mechanism from an activated to a barrierless dynamics is reflected in the plot of variation of the rate with supersaturation or density, as shown in **Fig. 7**. The change in the slope at the density (equivalent to supersaturation) $\rho_{ks}^*$ signifies the crossover from the activated to the barrierless growth. We call the density $\rho_{ks}^*$ as the onset density for the kinetic spinodal point **[32, 33]**. The calculated rate of nucleation scales differently with supersaturation below and above the kinetic spinodal. This kinetic spinodal is located at a temperature (or, density) lower (higher) than the thermodynamic spinodal point. At supersaturations between the kinetic and thermodynamic spinodal points, the clusters grow through *barrierless diffusion*. Just beyond the kinetic spinodal, only a few clusters grow whereas at the thermodynamic spinodal point all the clusters grow spontaneously.

Thus, we identify the disappearance of the free energy barrier for the largest liquid-like cluster as signifying the onset of the kinetic spinodal point which is the actual limit of metastability of the system. Our preliminary numerical calculations seem to support the previous analysis, for small system sizes. On the other hand, thermodynamic spinodal point corresponds to the disappearance of all $F_k^L(n^*)$, i.e. the point where $F(n^*)$ is zero, is not system size dependent and also is not relevant for most cases.

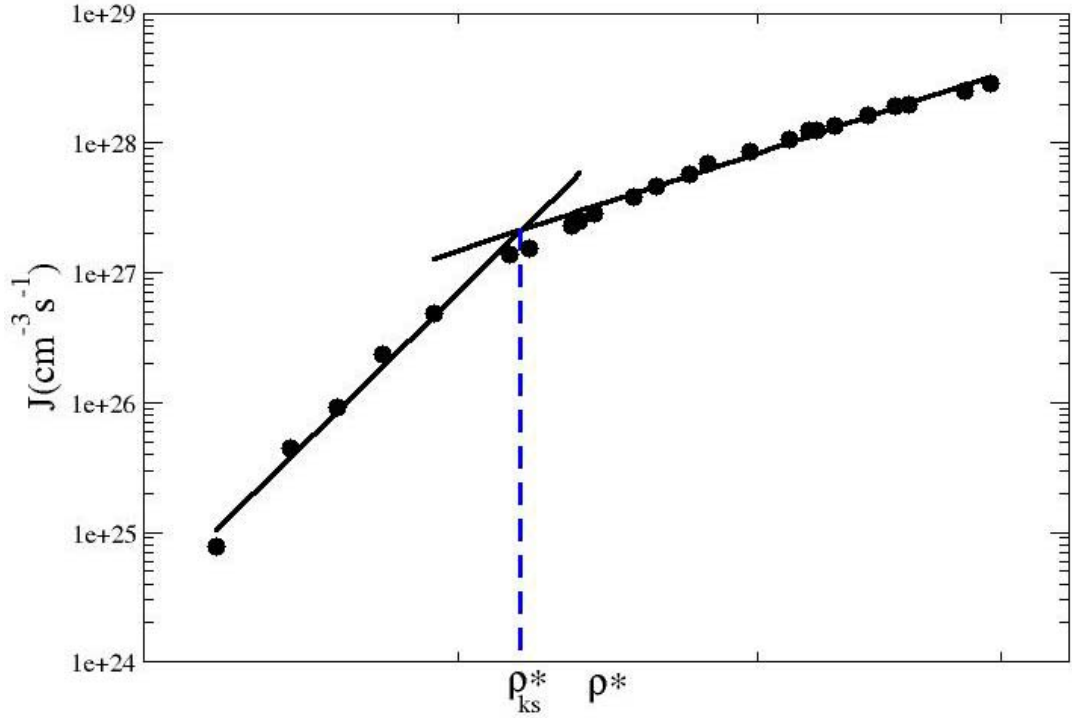

**Figure – 7. The rate of nucleation calculated from mean first passage time at different supersaturations at $T^*=0.741$ and $N = 500$. It clearly shows the crossover of nucleation from activated to barrierless diffusion at density $\rho_{ks}^*=0.04183$, indicated by a dashed bar. The supersaturation corresponding to the dashed line is the kinetic spinodal point [32].**

Clearly, the nucleation scenario is different at large supersaturation when the system is in the barrierless growth regime (See **Fig. 5a**) and the growth of the cluster can be modeled as *a diffusion on a flat free energy surface in cluster's size space*, bounded by an absorbing barrier at the critical cluster size, $n_c$ and approximately by a reflecting barrier at the free energy minimum of the *largest cluster size*, $n_M(N)$. $N$ is the total number of particles in the system. The corresponding mean first passage time (i.e. the growth time) is given by the following well-known expression [44],

$$\tau = \int_{n_M(N)}^{n_C} \frac{1}{D(y)} dy\, e^{\beta \Delta G(y)} \int_{n_M(N)}^{y} dz\, e^{-\beta \Delta G(z)}, \qquad (6)$$

where $D$ is the diffusion constant in cluster size ($n$) space (n) and is given by [45],

$$D(n) = \frac{2P}{\sqrt{2\pi m k_B T}} \left(\frac{36\pi}{\rho_l^2}\right)^{1/3} n^{2/3} = D_0\, n^{2/3}. \qquad (7)$$

Assume that the free energy surface to be flat (*i.e.* independent of cluster size) in between $n_M(n)$ and $n_C$, which is valid near kinetic spinodal point (see free energy of largest cluster in **Fig.5(a)** and **Fig.5(b)**). From **Eq. (6)** and **Eq. (7)** we obtain,

$$\tau = \frac{1}{D_0} \int_{n_M(N)}^{n_C} \frac{1}{y^{5/3}} dy\, (y - n_M(N)). \qquad (8)$$

Further integration gives rise to the following expression of $\tau$,

$$\tau = \frac{1}{D_0}\left[ 3\left(n_C^{1/3} - n_M^{1/3}(N)\right) - \frac{3}{2} n_M(N)\left(n_M^{-2/3}(N) - n_C^{-2/3}\right) \right]. \qquad (9)$$

After organizing the terms in **Eq. (9)** we obtain the following expression for the MFPT

$$\tau = \tau_0 \left[ \left(n_C^{1/3} - n_M^{1/3}(N)\right) + \frac{n_M(N)}{2}\left(n_C^{-2/3} - n_M^{-2/3}(N)\right) \right], \qquad (10)$$

where $\tau_0$ is the time constant associated to the diffusion constant in cluster size space ($n$) and is given by, $\tau_0 = \frac{3}{D_0} = \left(\frac{3\rho_l^2}{4\pi}\right)^{1/3} \frac{\sqrt{2\pi m k_B T}}{2P}$. $n_M(N)$ increases with $N$ while $n_C$ decreases with $S$.

Therefore, according to **Eq. (10)**, at constant supersaturation, the MFPT ($\tau$) decreases with $N$. For small clusters, as shown by **Fig. 5(a)**, $n_M(N)$ approaches to zero.

The snapshots of the trajectory of growth of clusters at two different supersaturation regimes (one far below and the other above the kinetic spinodal point) are shown in **Fig. 8(a) - 8(c)**. **Fig. 8(a)** depicts the situation at low supersaturation or below kinetic spinodal point. **Fig. 8(b)** and **Fig. 8(c)** depict situation at large supersaturation or above the kinetic spinodal point. From these snapshots it is clear that below the kinetic spinodal point, only one cluster crosses the critical nucleation barrier, although many other small clusters are present in the system. Above the kinetic spinodal, many clusters cross the critical size through barrierless diffusion.

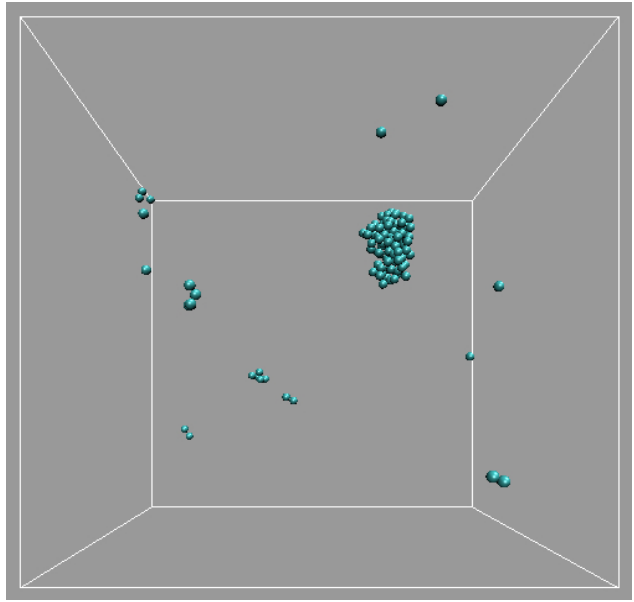

**Figure – 8(a). Snapshot of liquid-like particles in the system when the largest cluster crosses the critical size at low supersaturation of $\rho^*$=0.0334083 for $N$=4000. The size of the largest cluster in the figure is of one hundred liquid-like particles. From the snapshot it is clear that only the largest**

cluster undergoes nucleation and corresponding growth. Compared to the largest cluster all the other clusters are very small in size and are engulfed by the largest cluster.

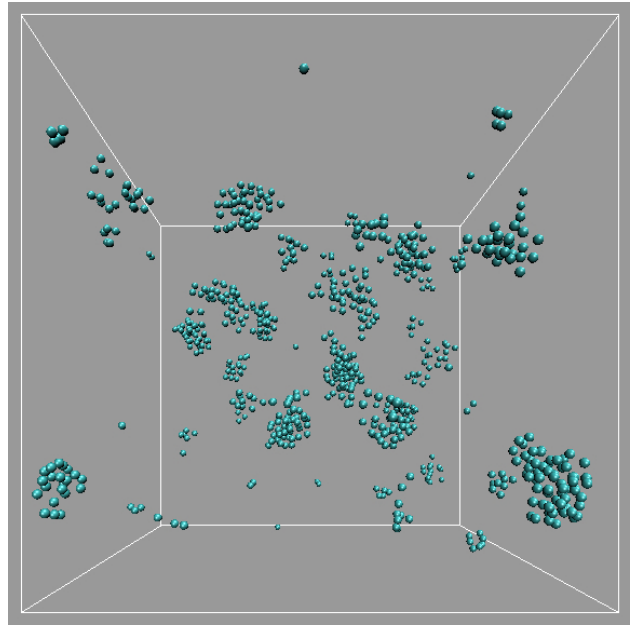

**Figure – 8(b).** The snapshot of all the liquid-like clusters present in the system when the largest cluster crosses the critical size at large supersaturation of $\rho^*$=0.05545 for *N*=4000. The size of the largest cluster in the figure is of one hundred liquid-like particles. Many liquid-like clusters are present in the system.

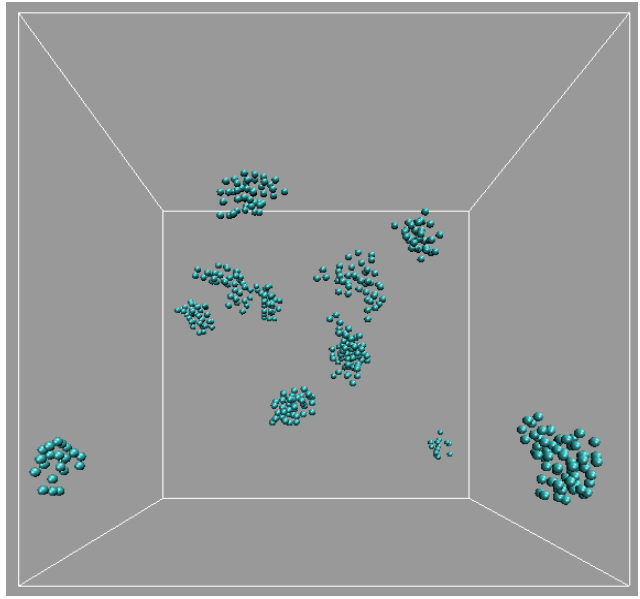

**Figure – 8(c). The snapshot of clusters in the system that have crossed the critical size at large supersaturation of $\rho^* = 0.05545$ for $N = 4000$. The size of the largest cluster in the figure is of one hundred liquid-like particles. The snapshot shows that at high supersaturation more than one cluster start nucleating and they are of comparable size.**

The classical nucleation theory models nucleation as a steady state flow of the gas (clusters below the critical size, mostly monomers and dimers) to the liquid (size of the cluster larger than the critical cluster) over an activation barrier, precisely the same way an activated chemical reaction is modeled. The well-known derivation of Zeldovich parallels Smoluchowski's well-known derivation of the rate of a chemical reaction in the over-damped limit. However, in nucleation, unlike in a chemical reaction, the conversion of a reactant to the product affects the system. The barrier crossing of the largest cluster leads to the immediate phase transformation as it engulfs the whole system. Therefore, in nucleation it is not meaningful to obtain the steady state rate at longer times. An earlier theoretical analysis of nucleation in Ising model systems found similar supersaturation dependence of the crossover of nucleation mechanism (from the activated to the barrierless) **[46]**.

## V. Nucleation in two dimensional systems

Two dimensional systems are important not only because of theoretical interest but also because of practical relevance. There are several 2D dimensional systems (though not perfectly 2D but can be considered to be) in nature. In material science thin film is a perfect example of 2D system. Physisorption of gases on solid surface **[47]**, surfactant monolayer adsorbed on an air/water interface **[48]**, thinning of completely wetted films on clean mica surfaces **[49]** can be considered as two dimensional fluids. In biological system transport of ions on cellular membrane is also in quasi-two dimensional space. Due to this diverse existence of two dimensional systems in nature there is a need of understanding the behavior of fluid in 2D. There are several theoretical, computational and experimental studies on liquid to solid phase transition. On the other hand there are very few studies on 2D gas-liquid phase transition despite the great importance of it **[50, 51]**. In this section we discuss about gas-liquid nucleation in 2D.

In two dimensions, the classical nucleation theory (CNT) expression for the free energy of the formation of a liquid nucleus of radius $R$ in grand canonical system is given by **[50, 51]**

$$\Delta\Omega(R) = -\pi R^2 \Delta P + 2\pi R \gamma_\infty \tag{11}$$

where $\Delta P = P_l - P_v$ is the pressure difference between the liquid-like droplet and the metastable gas phase, and $\gamma_\infty$ is the line tension of the planar interface. Here, the line tension for a nucleus of radius $R$ is assumed to be that of a planar surface between bulk phases using the capillary approximation. The maximum of $\Delta\Omega(R)$ gives the critical size of the nucleus ($R^*$) and the free energy barrier ($\Delta\Omega(R^*)$) as given by the following expression

$$R^* = \frac{\gamma_\infty}{\Delta P}, \qquad (12)$$

$$\Delta\Omega(R^*) = \frac{\pi\gamma_\infty^2}{\Delta P}. \qquad (13)$$

If the liquid is assumed to be incompressible and gas to be ideal, the pressure difference $\Delta P$ can be written in terms of $\Delta\mu = \mu - \mu_o$ or of the super-saturation $S = P_v/P_o$ as follows, where $\mu_o$ and $P_o$ are the chemical potential and pressure at gas-liquid coexistence, respectively.

$$\Delta P = \rho_l \Delta\mu = \rho_l k_B T \ln S \qquad (14)$$

Now as the activity $\xi$ is given as $\xi = exp(\mu/k_BT)/\lambda^2$, $S$ can be written in terms of $\xi$ as $S=\xi/\xi_0$. Here, $\lambda$ is the de Broglie wavelength and $\xi_0$ is the activity at the gas-liquid coexistence. The free energy barrier for the formation of the critical cluster is

$$\Delta\Omega^* = \frac{\pi\gamma_\infty^2}{\rho_l \Delta\mu} = \frac{\pi\gamma_\infty^2}{\rho_l k_B T \ln S}. \qquad (15)$$

The use of these expressions requires a priori knowledge of the free energy difference between the liquid and the vapor phase, and the line tension. While the former is available from the equation of state, it is somewhat harder to obtain a precise estimate of the line tension, although semi-quantitative estimates are often available.

The validity of CNT expression of the free energy barrier has been widely tested for gas-liquid nucleation in three dimensional (3D) systems, by carrying out both detailed theoretical and computer simulation studies. It is found that in 3D, the CNT free energy barrier overestimates the actual barrier by about 4-5 $k_BT$. However, such studies have hardly been carried out for 2D gas-liquid nucleation. Recently Santra *et al*. **[51]** studied 2D gas-liquid nucleation in great detail. Line tension is one of the key quantities in nucleation.

Santra *et al*. calculated the line tension using GC-TMMC **[52-55]** simulation for different interface length and extrapolated it to obtain the line tension in the thermodynamic limit following the finite size scaling method suggested by Binder **[56, 57]** (see **Fig.9**),

$$\gamma_L = c_1 \frac{1}{L} + c_2 \frac{\ln L}{L} + \gamma_\infty . \tag{16}$$

The line tension values for different cut-off are given in **Table I**. The line tension values for different potential cut-off radii shows that the growth of line tension with the cut-off values slows down and eventually converges as we go to the larger cut-off.

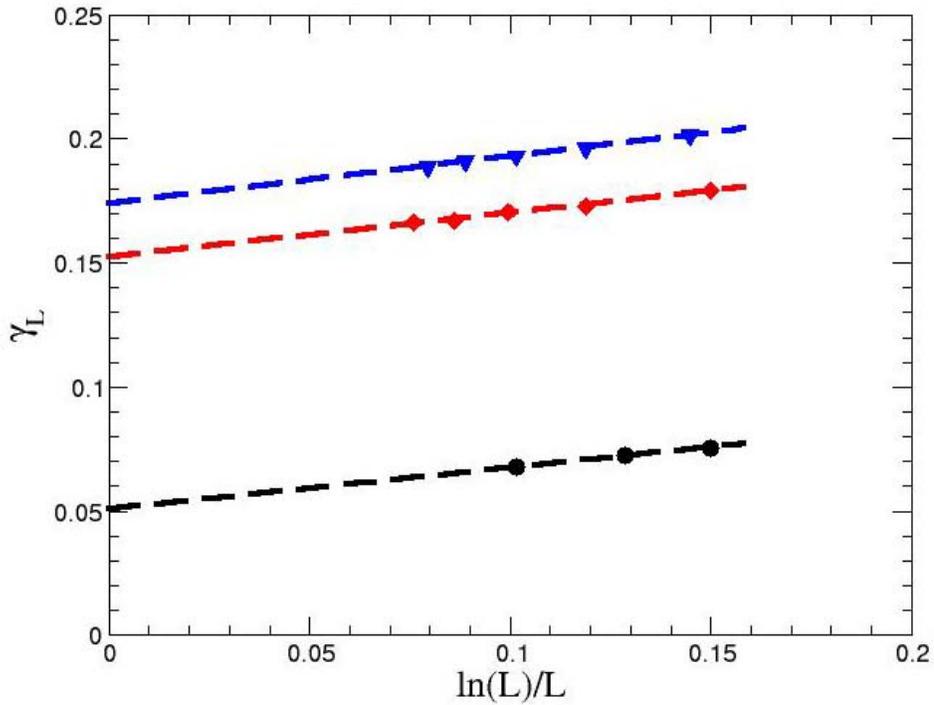

**Figure – 9.** The liquid-vapor interface energy per unit interface length vs. ln(L)/L where L is the length of the interface. In the figure circle, diamond and triangle-down are for $r_c$ =2.5σ, 4.0σ and 7.0σ, respectively. Dotted lines are the corresponding linear fits [51].

| $r_c$ | $\rho_v^*$ | $\rho_l^*$ | $\gamma^*$ | $n_{sim}^*$ | $\beta\Delta\Omega_{sim}^*$ | $n_{CNT}^*$ | $\beta\Delta\Omega_{CNT}^*$ | $J_{CNT}$ $(cm^{-2}s^{-1})$ | $J_{CNT}/J_{sim}$ |
|---|---|---|---|---|---|---|---|---|---|
| 2.5σ | 0.040 | 0.71 | 0.05(1) | 38 | 7.57 | 7 | 0.64 | $3.4\times10^7$ | 1022 |
| 4.0σ | 0.023 | 0.74 | 0.15(1) | 96 | 12.14 | 58 | 5.57 | $8.3\times10^4$ | 713 |
| 7.0σ | 0.023 | 0.74 | 0.17(3) | 104 | 13.84 | 77 | 7.31 | $1.4\times10^4$ | 685 |
| Full LJ | 0.023 | 0.74 | 0.17(8) | 104 | 14.22 | 81 | 7.34 | $1.4\times10^4$ | 972 |

**Table I.** The vapor-liquid line tension ($\gamma^*$), coexistence bulk densities of vapor ($\rho_v^*$), liquid ($\rho_l^*$) phases and nucleation data for 2D LJ system for different cut-off values. $n_{sim}^*$ and $n_{CNT}^*$ are the critical cluster size obtained from simulation and CNT, respectively. $J$'s are corresponding nucleation rate. $J_{CNT}$ has been calculated for Argon LJ system. Nucleation data are at saturation-ratio 1.1.

Earlier DFT calculation by Zeng **[50]** using temperature dependent diameter of particles finds that the surface tension for full *LJ* system at $T^* = 0.427$ is about 0.32. In recent computer simulation study it has been found to be much smaller ($\gamma^* = 0.178$). However, the critical temperature from DFT calculation ($T_c^* = 0.738$) is rather different from the simulation value ($T_c^* = 0.533$) **[58]**.

Bulk densities of liquid and vapor phases at coexistence for different cut-off potential are also reported in **Table I**. The density profile for equilibrium liquid-vapor interface is shown in **Fig. 10**. The densities of liquid and vapor phases change in going from potential cut-off value 2.5σ to 4.0σ. After that it remains unchanged for higher cut-off value.

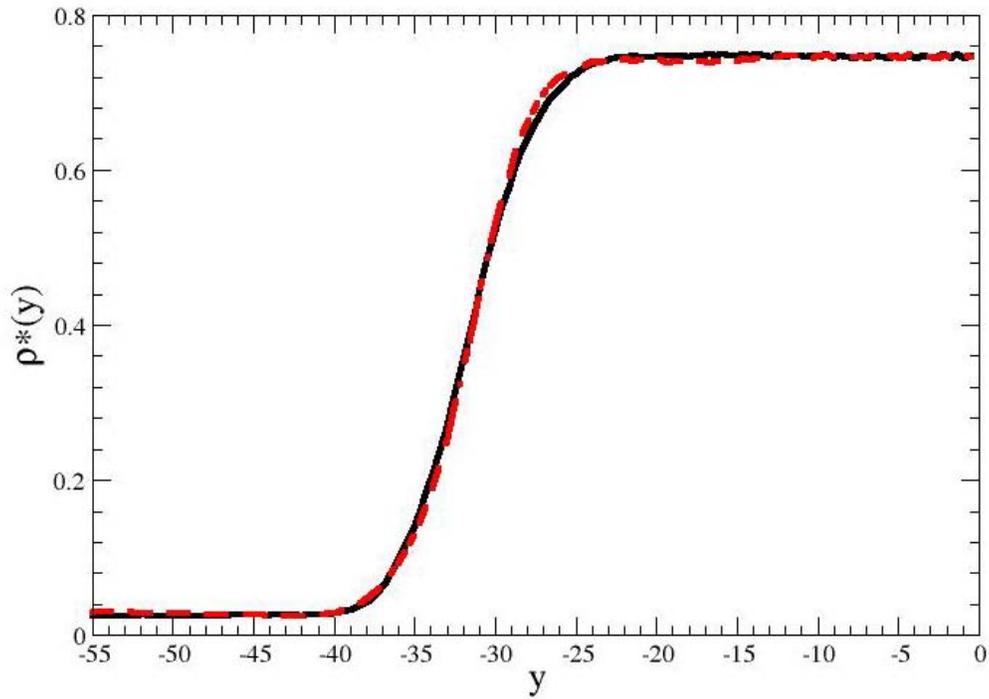

**Figure – 10. Density profile of equilibrium liquid-vapor interface $\rho^*(y)$ for a system with square box of length $L$ = 113σ at $T^*$ = 0.427, where the y-axis is perpendicular to the interface. A liquid-slab with two parallel liquid-vapor interfaces is created. Here, we have shown only one interface. Solid black line is for $r_c$ = 4.0σ and red dashed line is for $r_c$ = 7.0σ.**

Both the nucleation barrier and the size of the critical cluster increase with an increase in the cut-off radius (see **Fig. 11**). This is expected as the surface tension increases markedly as we increase the cut-off radius. It has been shown before that in 2D LJ system, the phase diagram is strongly dependent on the potential cut-off **[59]**. Now it shows that the nucleation phenomenon is also strongly dependent on the cut-off used. However, beyond the cut-off radius 7.0σ, there is only a negligible effect of cut-off on nucleation, and the converged result is obtained.

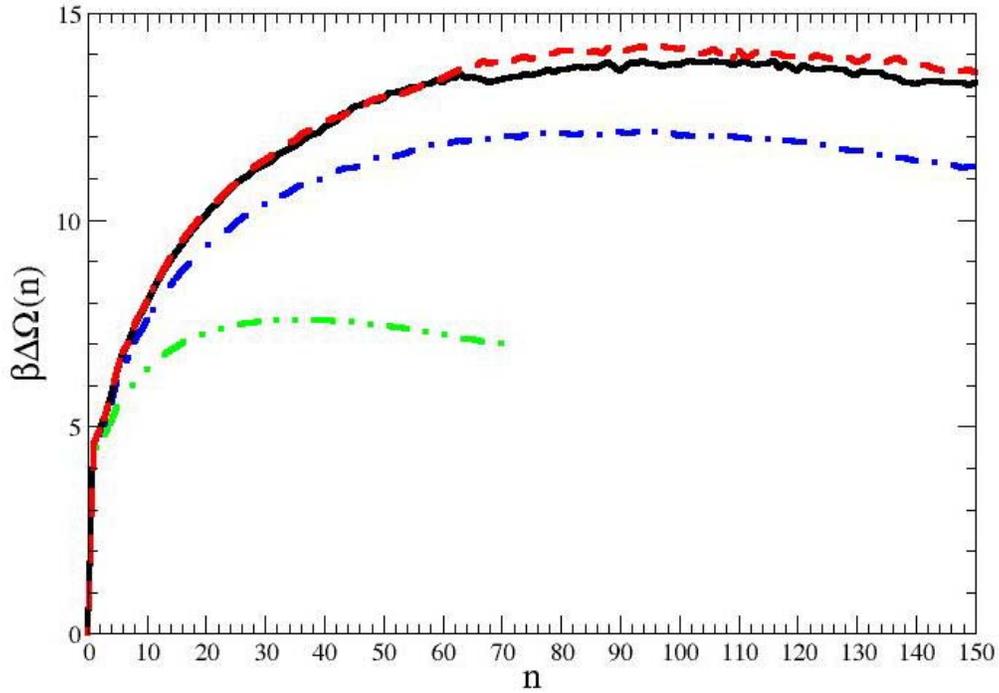

**Figure – 11. The computed free energy of formation of a liquid-like embryo as a function of the size of the embryo. Here *n* is the number of atoms in the cluster. Saturation-ratio *S* =1.1, $T^*$ = 0.427. Black solid line is for $r_c$ = 7.0σ, blue dot-dashed line for $r_c$ = 4.0σ, green dot-dot-dashed line for $r_c$ = 2.5σ and red long-dashed line for full LJ potential [51].**

Comparison of the free energy barriers obtained from simulation with those predicted by CNT (with the line tension obtained from simulation) at saturation-ratio $S = 1.1$ indicates that CNT *significantly underestimates* the nucleation barrier for all cut-off at $T^* = 0.427$ [see **Table I**]. Also, the critical nucleus size predicted by CNT is much smaller than the observed value. The discrepancy remains even without cutoff. In fact, the free energy barrier prediction of CNT is found *not to be accurate* for any of the cut-off at $S = 1.1$.

The rate of nucleation predicted by CNT is,

$$J = J_0 e^{-\Delta\Omega^*/k_B T}.  \qquad (17)$$

where, $J_0$ is the pre-factor used in CNT and is given as **[50]**,

$$J_0 = \frac{\rho_v^2}{S}\sqrt{\frac{\Delta\mu a_1}{2\pi m}}. \qquad (18)$$

where $m$ is the atomic mass, $\rho_v$ is the number density of gas, and $a_1$ is the area per particle in the bulk liquid. The ratio $J_{CNT}/J_{sim}$ is shown in **Table I** for different cut-off values. $J_{CNT}$ *overestimates* $J_{sim}$ by more than two orders of magnitude, for all of the cut-off values.

Interestingly, while CNT overestimates the barrier in 3D system, it underestimates the free energy barrier in 2D! However, in the case of 3D nucleation for a range of saturation-ratio starting from $S = 1.53$ to $2.24$, there is a relatively small, nearly constant offset of nucleation free energy barrier ($\sim 5 k_B T$) between CNT prediction and computer simulation result. A recent study on 3D LJ system shows that this discrepancy is much higher for smaller clusters **[60]**. In 2D, even close to the coexistence, the CNT underestimates the nucleation barrier by $\sim 6.5\, k_B T$ which is about 50% smaller than the correct value. Whereas in case of 3D nucleation the critical cluster size prediction of CNT is almost accurate, in 2D system the prediction is largely deviated from simulation value, e.g. in case of $r_c = 7.0\sigma$ at $S = 1.1$ CNT underestimates the critical cluster size

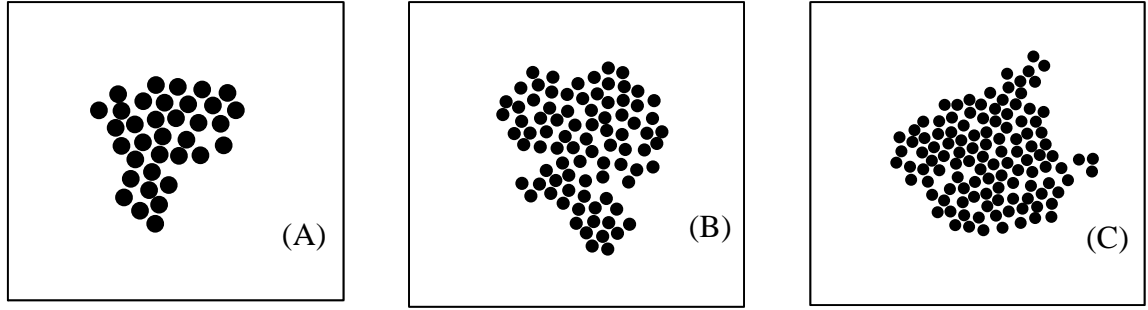

**Figure – 12. Snapshots of critical liquid-like clusters at $S = 1.1$, $T^* = 0.427$ with the following cut-off radii. (A) $r_c = 2.5\sigma$, size of the critical cluster $n^* = 35$. (B) $r_c = 4.0\sigma$, $n^* = 85$. (C) $r_c = 7.0\sigma$, $n^* = 115$.**

The snapshots of critical clusters show that they are deviated from circular shape (**Fig.12**). For large clusters voids are present inside the liquid clusters, but such voids are present even in bulk liquid phase in 2D due to large scale thermal fluctuations.

It has been found in case of crystallization in 3D LJ system that if the nuclei are assumed ellipsoidal in shape instead of spherical, the nucleation free energy barrier given by CNT is improved considerably **[30]**. It also improves the size of the critical cluster. The attempt of similar shape correction in 2D improves the free energy barrier in small amount but the critical cluster size gets improved a lot and it almost matches with the simulation value (**Fig.13**). Figure 13 and the accompanied analysis discussed above clearly shows the importance of non-circular shape of the clusters in determining the free energy barrier of nucleation. The length of the periphery using a more accurate method gives far better improvement in the free energy barrier than the ellipse assumption and it is close to the simulated free energy barrier.

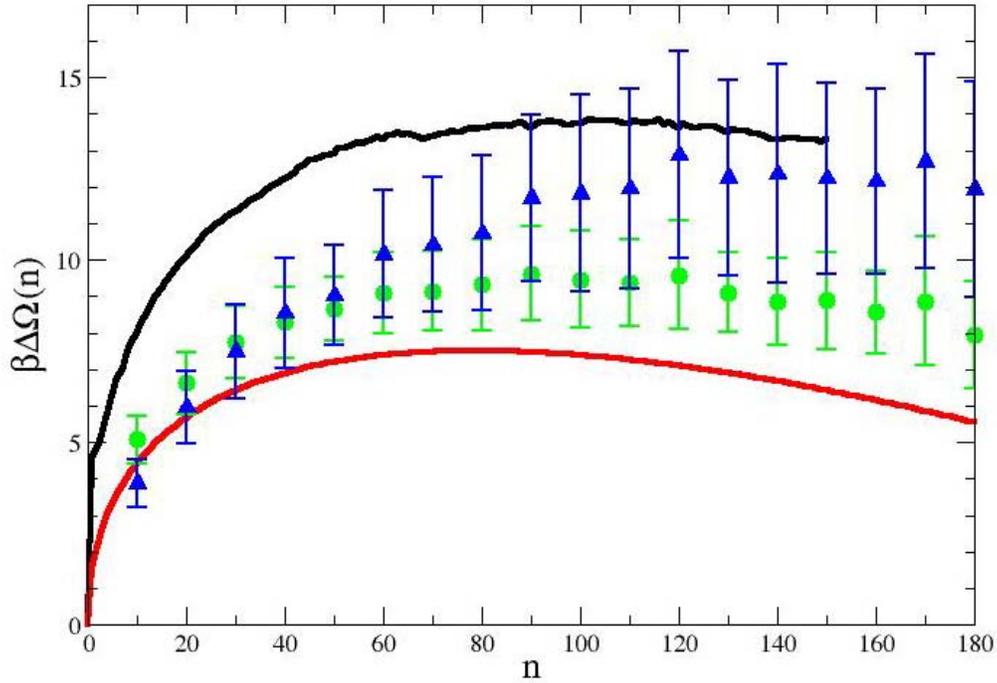

**Figure – 13. The free energy of nucleation versus size of a liquid cluster. Black line is from simulation, red line is CNT calculation, Circle with error bar is after shape correction approximating shape of the cluster as an ellipse. Triangle up with error bar represents shape correction from direct perimeter calculation. Here $T^* = 0.427$, $r_c = 7.0$, $S = 1.1$ [51].**

## VI. Effects of the range of interaction potential on Gas-liquid nucleation

The range of interaction potential in simple fluid plays a significant role on determining the structure and properties of gas-liquid interface. The above and also the majority of computational and theoretical studies of gas-liquid and liquid-solid nucleation have been carried out with a Lennard-Jones potential which is truncated at 2.5 times the molecular diameter (that is, the Lennard-Jones parameter σ). The potential is then shifted to preserve the continuity of the potential. In the high density liquid phase, this truncation procedure does not seem to introduce any serious error, and the effects of increasing the range of interaction are estimated to be negligible. The situation turned out to be quite different in the case of nucleation **[51, 61]**.

Many computational and theoretical studies show that the surface tension of gas-liquid interface increases by sufficiently large amount on going from small to high cut-off **[61-64]**. This dependence of surface tension is reflected to both the nucleation barrier and the size of the critical cluster. The high sensitivity of the line tension, nucleation free energy barrier and the critical cluster size to the range of interaction was systematically demonstrated in the case of two dimensional gas-liquid nucleation **[51]**. Later the same was observed, to somewhat lesser extent, for three dimensional systems **[61]**. The free energy barrier of nucleation increases strongly with increasing cut-off and the limit of metastability (spinodal) also increases. Though the values of the properties associated to nucleation change quantitatively with the range of interaction potential the results remain same qualitatively.

Fortunately, however, the quantitative aspects did not undergo too much of change. The main change was that the kinetic spinodal point got shifted to higher value of supersaturation *S*. The new features discussed here, like the simultaneous growth of many clusters near the kinetic spinodal limit remained unchanged. The crossover behavior from activated to barrierless growth of clusters also remained unaltered. Actually, the sensitivity of the surface tension and the nucleation parameters to the range of interaction could be exploited to gain additional insight into the nucleation mechanism at large metastability.

## VII. Conclusion

In this article we have summarized results of the studies on nucleation at large metastability, in both two and three dimensional systems. In the limit of large metastability the mechanism of nucleation becomes collective and diffuse over the system. As many clusters grow simultaneously, the classical nucleation theory with only one cluster can no longer faithfully describe the nucleation scenario.

A new set of order parameters have been introduced to describe the collective nature of nucleation at large metastability. These new set of parameters quantify the growth of clusters ordered according to their size. Thus, k=1 corresponds to the largest cluster, while k=2 as the second largest cluster and so on. We then present both analytical and numerical free energy surface of these clusters and also in simulation studied time evolution and growth of these clusters.

These studies seem to provide a unified picture of nucleation at large metastability which can be articulated as follows. At low metastability, the free energy of the largest cluster coincides with the CNT free energy, both having large free energy barrier. In simulation we find under the same conditions, only one cluster grows. The picture at large metastability is different. Now, the barrier for growth of several largest clusters becomes very small (of the order of $k_BT$). The simulation studies show that many clusters grow simultaneously. Since the free energy surface is flat, the growth in the cluster size space is via diffusion on a flat free energy surface. Since the large but sub-critical clusters form at a relatively small span of time after the quench of temperature or pressure, the main determinant of nucleation at large metastability is this diffusive growth.

The consequence of this change in mechanism of growth is reflected in a change in the supersaturation dependence of the nucleation rate --- the supersaturation dependence becomes weaker at large metastability because of transition from activated to barrierless regime. This is captured both in free energy calculations with the extended set of order parameters and in simulations. ***The classical nucleation theory, however, fails to describe this crossover at large metastability.*** This we believe is one of the clearest demonstrations of the breakdown of CNT.

The dependence of surface tension on the range of interaction has been found surprising strong, both for two and three dimensional systems. The free energy of the liquid state also decreases as the range of interaction increases. Thus, the quantitative details of nucleation studies in simulations undergo substantial change, However, the qualitative features at large metastability remains essentially the same.

## Acknowledgement

We thank Prof. B. Prabhakar and Dr. S. Chakraborty for collaboration at the early stage of the project. We also thank Mr. B. Jana for discussions. This work was supported in parts by grants from DST and CSIR (India). BB thanks DST for a Sir JC Bose Fellowship.